\documentclass[10pt,a4paper,onecolumn]{article}
\usepackage{marginnote}
\usepackage{graphicx}
\usepackage{xcolor}
\usepackage{authblk,etoolbox}
\usepackage{titlesec}
\usepackage{calc}
\usepackage{tikz}
\usepackage{hyperref}
\hypersetup{colorlinks,breaklinks,
            urlcolor=[rgb]{0.0, 0.5, 1.0},
            linkcolor=[rgb]{0.0, 0.5, 1.0}}
\usepackage{caption}
\usepackage{tcolorbox}
\usepackage{amssymb,amsmath}
\usepackage{ifxetex,ifluatex}
\usepackage{seqsplit}
\usepackage{xstring}

\usepackage{float}
\let\origfigure\figure
\let\endorigfigure\endfigure
\renewenvironment{figure}[1][2] {
    \expandafter\origfigure\expandafter[H]
} {
    \endorigfigure
}

\usepackage{fixltx2e} 
\usepackage[
  backend=biber,
]{biblatex}
\bibliography{paper.bib}

\let\textttOrig=\texttt
\def\texttt#1{\expandafter\textttOrig{\seqsplit{#1}}}
\renewcommand{\seqinsert}{\ifmmode
  \allowbreak
  \else\penalty6000\hspace{0pt plus 0.02em}\fi}

\makeatletter
\let\href@Orig=\href
\def\href@Urllike#1#2{\href@Orig{#1}{\begingroup
    \def\Url@String{#2}\Url@FormatString
    \endgroup}}
\def\href@Notdoi#1#2{\def\tempa{#1}\def\tempb{#2}%
  \ifx\tempa\tempb\relax\href@Urllike{#1}{#2}\else
  \href@Orig{#1}{#2}\fi}
\def\href#1#2{%
  \IfBeginWith{#1}{https://doi.org}%
  {\href@Urllike{#1}{#2}}{\href@Notdoi{#1}{#2}}}
\makeatother

\usepackage[top=3.5cm, bottom=3cm, right=1.5cm, left=1.0cm,
            headheight=2.2cm, reversemp, includemp, marginparwidth=4.5cm]{geometry}

\titleformat{\section}
  {\normalfont\sffamily\Large\bfseries}
  {}{0pt}{}
\titleformat{\subsection}
  {\normalfont\sffamily\large\bfseries}
  {}{0pt}{}
\titleformat{\subsubsection}
  {\normalfont\sffamily\bfseries}
  {}{0pt}{}
\titleformat*{\paragraph}
  {\sffamily\normalsize}

\usepackage{fancyhdr}
\makeatletter
\let\ps@plain\ps@fancy
\fancyheadoffset[L]{4.5cm}
\fancyfootoffset[L]{4.5cm}

\definecolor{linky}{rgb}{0.0, 0.5, 1.0}

\newtcolorbox{repobox}
   {colback=red, colframe=red!75!black,
     boxrule=0.5pt, arc=2pt, left=6pt, right=6pt, top=3pt, bottom=3pt}

\newcommand{\ExternalLink}{%
   \tikz[x=1.2ex, y=1.2ex, baseline=-0.05ex]{%
       \begin{scope}[x=1ex, y=1ex]
           \clip (-0.1,-0.1)
               --++ (-0, 1.2)
               --++ (0.6, 0)
               --++ (0, -0.6)
               --++ (0.6, 0)
               --++ (0, -1);
           \path[draw,
               line width = 0.5,
               rounded corners=0.5]
               (0,0) rectangle (1,1);
       \end{scope}
       \path[draw, line width = 0.5] (0.5, 0.5)
           -- (1, 1);
       \path[draw, line width = 0.5] (0.6, 1)
           -- (1, 1) -- (1, 0.6);
       }
   }

\patchcmd{\@maketitle}{center}{flushleft}{}{}
\patchcmd{\@maketitle}{center}{flushleft}{}{}
\patchcmd{\@maketitle}{\LARGE}{\LARGE\sffamily}{}{}
\def\maketitle{{%
  
  \AB@maketitle}}
\makeatletter
\renewcommand\AB@affilsepx{ \protect\Affilfont}
\renewcommand\AB@affilnote[1]{{\bfseries #1}\hspace{3pt}}
\renewcommand{\affil}[2][]%
   {\newaffiltrue\let\AB@blk@and\AB@pand
      \if\relax#1\relax\def\AB@note{\AB@thenote}\else\def\AB@note{#1}%
        \setcounter{Maxaffil}{0}\fi
        \begingroup
        \let\href=\href@Orig
        \let\texttt=\textttOrig
        \let\protect\@unexpandable@protect
        \def\thanks{\protect\thanks}\def\footnote{\protect\footnote}%
        \@temptokena=\expandafter{\AB@authors}%
        {\def\\{\protect\\\protect\Affilfont}\xdef\AB@temp{#2}}%
         \xdef\AB@authors{\the\@temptokena\AB@las\AB@au@str
         \protect\\[\affilsep]\protect\Affilfont\AB@temp}%
         \gdef\AB@las{}\gdef\AB@au@str{}%
        {\def\\{, \ignorespaces}\xdef\AB@temp{#2}}%
        \@temptokena=\expandafter{\AB@affillist}%
        \xdef\AB@affillist{\the\@temptokena \AB@affilsep
          \AB@affilnote{\AB@note}\protect\Affilfont\AB@temp}%
      \endgroup
       \let\AB@affilsep\AB@affilsepx
}
\makeatother

\renewcommand\Affilfont{\sffamily\small\mdseries}
\ifnum 0\ifxetex 1\fi\ifluatex 1\fi=0 
  \usepackage[T1]{fontenc}
  \usepackage[utf8]{inputenc}

\else 
  \ifxetex
    \usepackage{mathspec}
  \else
    \usepackage{fontspec}
  \fi
  \defaultfontfeatures{Ligatures=TeX,Scale=MatchLowercase}

\fi
\IfFileExists{upquote.sty}{\usepackage{upquote}}{}
\IfFileExists{microtype.sty}{%
\usepackage{microtype}
\UseMicrotypeSet[protrusion]{basicmath} 
}{}

\usepackage{hyperref}
\hypersetup{unicode=true,
            pdftitle={MiSTree: a Python package for constructing and analysing Minimum Spanning Trees},
            pdfborder={0 0 0},
            breaklinks=true}
\let\addcontentslineOrig=\addcontentsline
\def\addcontentsline#1#2#3{\bgroup
  \let\texttt=\textttOrig\addcontentslineOrig{#1}{#2}{#3}\egroup}
\let\markbothOrig\markboth
\def\markboth#1#2{\bgroup
  \let\texttt=\textttOrig\markbothOrig{#1}{#2}\egroup}
\let\markrightOrig\markright
\def\markright#1{\bgroup
  \let\texttt=\textttOrig\markrightOrig{#1}\egroup}

\usepackage{graphicx,grffile}
\makeatletter
\def\maxwidth{\ifdim\Gin@nat@width>\linewidth\linewidth\else\Gin@nat@width\fi}
\def\maxheight{\ifdim\Gin@nat@height>\textheight\textheight\else\Gin@nat@height\fi}
\makeatother
\setkeys{Gin}{width=\maxwidth,height=\maxheight,keepaspectratio}
\IfFileExists{parskip.sty}{%
\usepackage{parskip}
}{
\setlength{\parindent}{0pt}
\setlength{\parskip}{6pt plus 2pt minus 1pt}
}
\providecommand{\tightlist}{%
  \setlength{\itemsep}{0pt}\setlength{\parskip}{0pt}}
\ifx\paragraph\undefined\else
\let\oldparagraph\paragraph
\renewcommand{\paragraph}[1]{\oldparagraph{#1}\mbox{}}
\fi
\ifx\subparagraph\undefined\else
\let\oldsubparagraph\subparagraph
\renewcommand{\subparagraph}[1]{\oldsubparagraph{#1}\mbox{}}
\fi

\title{MiSTree: a Python package for constructing and analysing Minimum
Spanning Trees}

        \author[1]{Krishna Naidoo}
    
      \affil[1]{Department of Physics \& Astronomy, University College London, Gower
Street, London, WC1E 6BT, UK}
  \date{\vspace{-5ex}}

\begin{document}
\maketitle

\marginpar{
  \sffamily\small

  {\bfseries DOI:} \href{https://doi.org/10.21105/joss.01721}{\color{linky}{10.21105/joss.01721}}

  \vspace{2mm}

  {\bfseries Software}
  \begin{itemize}
    \setlength\itemsep{0em}
    \item \href{https://github.com/openjournals/joss-reviews/issues/1721}{\color{linky}{Review}} \ExternalLink
    \item \href{https://github.com/knaidoo29/mistree}{\color{linky}{Repository}} \ExternalLink
    \item \href{http://doi.org/10.5281/zenodo.3495008}{\color{linky}{Archive}} \ExternalLink
  \end{itemize}

  \vspace{2mm}

  {\bfseries Submitted:} 27 August 2019\\
  {\bfseries Published:} 17 October 2019

  \vspace{2mm}
  {\bfseries License}\\
  Authors of papers retain copyright and release the work under a Creative Commons Attribution 4.0 International License (\href{https://creativecommons.org/licenses/by/4.0/}{\color{linky}{CC-BY}}).
}

\section{Summary}\label{summary}

The \emph{minimum spanning tree} (MST), a graph constructed from a
distribution of points, draws lines between pairs of points so that all
points are linked in a single skeletal structure that contains no loops
and has minimal total edge length. The MST has been used in a broad
range of scientific fields such as particle physics (to distinguish
classes of events in collider collisions, see Rainbolt and Schmitt
(2017)), in astronomy (to detect mass segregation in star clusters, see
Allison et al. (2009)) and cosmology (to search for filaments in the
cosmic web, see Alpaslan et al. (2014)). Its success in these fields has
been driven by its sensitivity to the spatial distribution of points and
the patterns within. \emph{MiSTree}, a public \emph{Python} package,
allows a user to construct the MST in a variety of coordinates systems,
including Celestial coordinates used in astronomy. The package enables
the MST to be constructed quickly by initially using a \(k\)-nearest
neighbour graph (\(k\)NN, rather than a matrix of pairwise distances)
which is then fed to Kruskal's algorithm (Kruskal 1956) to construct the
MST. \emph{MiSTree} enables a user to measure the statistics of the MST
and provides classes for binning the MST statistics (into histograms)
and plotting the distributions. Applying the MST will enable the
inclusion of high-order statistics information from the cosmic web which
can provide additional information to improve cosmological parameter
constraints (Naidoo et al. 2019). This information has not been fully
exploited due to the computational cost of calculating \(N\)-point
statistics. \emph{MiSTree} was designed to be used in cosmology but
could be used in any field which requires extracting non-Gaussian
information from point distributions.

\section{Motivation}\label{motivation}

Studies of point distributions often measure their 2-point statistics
(i.e.~the distribution of distances between pairs of points) which are
then compared to theoretical models. This is a powerful technique and
has been used very successfully in the field of cosmology to study the
early Universe and the large scale distribution of galaxies.
Unfortunately this statistic can only fully describe a distribution that
is Gaussian, if it is non-Gaussian then the 2-point is no longer
sufficient. The conventional method to incorporate non-Gaussian
information is to look at the distribution's \(N\)-point statistic (if
\(N\!=\!3\) we look at the distribution of triangles, if \(N\!=\!4\) we
look at the distribution of quadrilaterals and so on). This method is
well motivated as in principle all the information that can describe a
distribution of points is contained within its \(N\)-point statistics
(see Szapudi and Szalay (1998)). However, calculating \(N\)-point
statistics even for \(N\!>\!3\) becomes quickly intractable for large
data sets.

The MST offers an alternative approach; the MST graph draws lines
between pairs of points so that all points are linked in a single
skeletal structure that contains no loops and has minimal total edge
length. Unlike \(N\)-point statistics, that typically scale by
\(\mathcal{O}(n^{N})\) for \(n\) points, the MST (computed using the
Kruskal algorithm (Kruskal 1956) which sequentially adds edges, from
shortest to longest, with the condition that the added edge does not
form a loop) can be constructed much faster (at best
\(\mathcal{O}(n\log n)\)). While the MST does not contain all the
information present in \(N\)-point statistics, it enables some of this
information to be captured and allows the identification of skeletal
patterns, as such it has found a broad range of applications in physics:
such as finding filaments in the distribution of galaxies (Alpaslan et
al. 2014), classifying particle physics collisions (Rainbolt and Schmitt
2017) and mass segregation in star clusters (Allison et al. 2009). The
MST has also been used in a number of other scientific field such as
computer science, sociology and epidemiology.

While algorithms to construct the \emph{minimum spanning tree} are well
known (e.g. Prim (1957) and Kruskal (1956)) implementations of these
often require the input of a matrix of pairwise distances. For a large
data set the creation of this matrix (with \(n^{2}\) elements) can be a
significant strain on memory while also making the construction of the
MST slower (\(\mathcal{O}(n^{2}\log n)\)).

\begin{figure}
\centering
\includegraphics{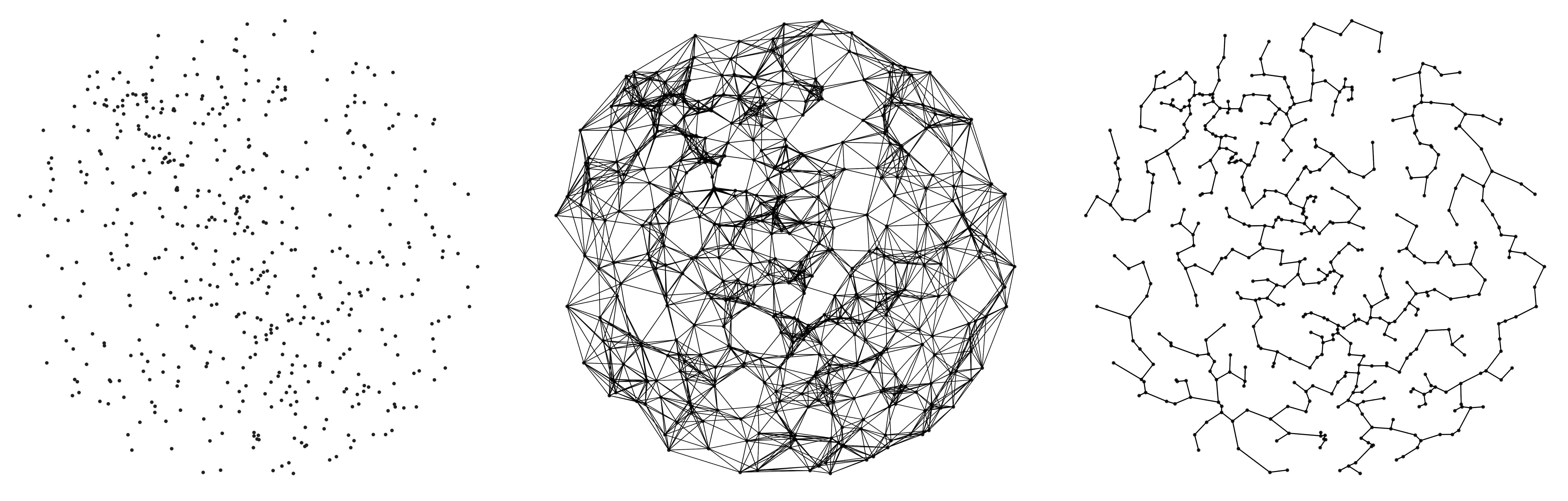}
\caption{An example of how \emph{MiSTree} constructs the MST from a
distribution of points (shown on the left). \emph{MiSTree} first begins
by constructing a \(k\)NN graph which links all points to their nearest
\(k\) neighbours (shown in the centre) and then runs the Kruskal
algorithm to construct the MST (shown on the right).}
\end{figure}

\section{MiSTree}\label{mistree}

\emph{MiSTree} is a public \emph{Python} package for the construction
and analysis of the MST. The package initially creates a \(k\)-nearest
neighbour graph (\(k\)NN, a graph that links each point to the nearest
\(k\) neighbours, using \emph{scikit-learn}'s
\texttt{kneighbours\_graph} function) which improves speed by limiting
the number of considered edges from \(n^{2}\) to \(kn\) (where
\(k\!\ll\! n\)) and then runs the Kruskal algorithm (Kruskal 1956)
(using \emph{scipy}'s \texttt{minimum\_spanning\_tree} function). The
stages of the MST construction are shown in Figure 1.

The MST can be constructed from data provided in 2/3 dimensions and in
tomographic (on a unit sphere) or spherical polar coordinates. The
weights of the edges are assumed to be the distances between points;
i.e.~the Euclidean distance for 2/3 dimension and spherical polar
coordinates, and angular distances for tomographic coordinates.
Furthermore, the package can very quickly measure the standard
statistics:

\begin{itemize}
\tightlist
\item
  degree (\(d\)) -- the number of edges attached to each node.
\item
  edge length (\(l\)) -- the length of edges in the MST.
\end{itemize}

While also being able to measure the statistics of branches, which are
defined as chains of edges connected with degree \(=2\):

\begin{itemize}
\tightlist
\item
  branch length (\(b\)) -- the sum of the lengths of member edges.
\item
  branch shape (\(s\)) -- the straight line distance between the tips of
  branches divided by the branch length.
\end{itemize}

The statistics calculated by \emph{MiSTree} are extensively explored in
Naidoo et al. (2019) and found to significantly improve constraints on
cosmological parameters when tested on simulations.

\begin{figure}
\centering
\includegraphics{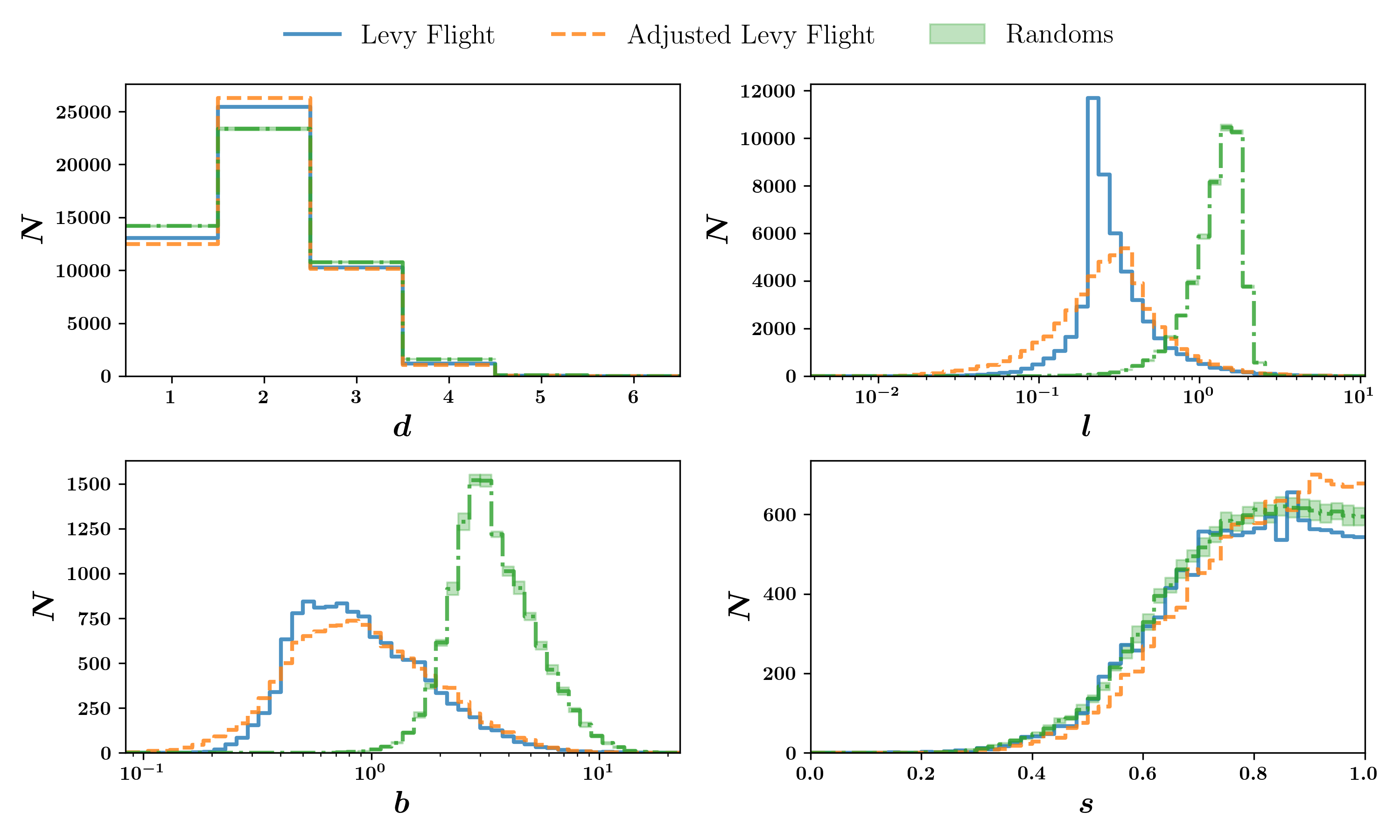}
\caption{Histograms of the distribution of the MST statistics degree
(\(d\)), edge length (\(l\)), branch length (\(b\)) and branch shape
(\(s\)) for a Levy Flight and Adjusted Levy Flight distribution in
comparison to a set of random distribution (details of which are
provided in Naidoo et al. 2019) in 3 dimensions.}
\end{figure}

\section{Basic Usage}\label{basic-usage}

To construct the MST using \emph{MiSTree} from a distribution of points
in 2 dimensions you would use the following commands:

\begin{verbatim}
import mistree as mist

# initialise MiSTree Minimum Spanning Tree class
mst = mist.GetMST(x=x, y=y)
mst.construct_mst()
\end{verbatim}

Once the MST is constructed it can either be used to look for features
in the distribution or to measure statistics of the graph which in turn
tell us about how points have been distributed. \emph{MiSTree} can
measure four statistics by default, which can be calculated directly
after initialising the \texttt{GetMST} class (an example of the
distribution of these statistics is shown in Figure 2):

\begin{verbatim}
d, l, b, s = mst.get_stats()
\end{verbatim}

The source code can be found on
\href{https://github.com/knaidoo29/mistree}{github} while documentation
and more complicated tutorials are provided
\href{https://knaidoo29.github.io/mistreedoc/}{here}.

\section{Dependencies}\label{dependencies}

Dependencies for \emph{MiSTree} include the \emph{Python} modules
\emph{numpy} (Oliphant 2006), \emph{matplotlib} (Hunter 2007),
\emph{scipy} (Jones et al. 2001), \emph{scikit-learn} (Pedregosa et al.
2011) and \emph{f2py} (Peterson 2009) (the latter of which is used to
compile \texttt{Fortran} subroutines).

\section{Acknowledgement}\label{acknowledgement}

I thank Ofer Lahav and Lorne Whiteway for their guidance and suggestions
in developing this package and acknowledge support from the Science and
Technology Facilities Council grant ST/N50449X.

\section*{References}\label{references}
\addcontentsline{toc}{section}{References}

\hypertarget{refs}{}
\hypertarget{ref-allison:2009}{}
Allison, R. J., S. P. Goodwin, R. J. Parker, S. F. Portegies Zwart, R.
de Grijs, and M. B. N. Kouwenhoven. 2009. ``Using the minimum spanning
tree to trace mass segregation'' 395 (May): 1449--54.
doi:\href{https://doi.org/10.1111/j.1365-2966.2009.14508.x}{10.1111/j.1365-2966.2009.14508.x}.

\hypertarget{ref-gama:2014}{}
Alpaslan, M., A. S. G. Robotham, S. Driver, P. Norberg, I. Baldry, A. E.
Bauer, J. Bland-Hawthorn, et al. 2014. ``Galaxy And Mass Assembly
(GAMA): the large-scale structure of galaxies and comparison to mock
universes'' 438 (February): 177--94.
doi:\href{https://doi.org/10.1093/mnras/stt2136}{10.1093/mnras/stt2136}.

\hypertarget{ref-Hunter:2007}{}
Hunter, J. D. 2007. ``Matplotlib: A 2d Graphics Environment.''
\emph{Computing in Science \& Engineering} 9 (3). IEEE COMPUTER SOC:
90--95.
doi:\href{https://doi.org/10.1109/MCSE.2007.55}{10.1109/MCSE.2007.55}.

\hypertarget{ref-scipy}{}
Jones, Eric, Travis Oliphant, Pearu Peterson, and others. 2001. ``SciPy:
Open Source Scientific Tools for Python.'' \url{http://www.scipy.org/}.

\hypertarget{ref-kruskal1956shortest}{}
Kruskal, Joseph B. 1956. ``On the Shortest Spanning Subtree of a Graph
and the Traveling Salesman Problem.'' \emph{Proceedings of the American
Mathematical Society} 7 (1). JSTOR: 48--50.
doi:\href{https://doi.org/10.2307/2033241}{10.2307/2033241}.

\hypertarget{ref-naidoo:2019}{}
Naidoo, Krishna, Lorne Whiteway, Elena Massara, Davide Gualdi, Ofer
Lahav, Matteo Viel, Héctor Gil-Marín, and Andreu Font-Ribera. 2019.
``Beyond two-point statistics: using the Minimum Spanning Tree as a tool
for cosmology.'' \emph{arXiv E-Prints}, July, arXiv:1907.00989.

\hypertarget{ref-numpy}{}
Oliphant, Travis. 2006. ``NumPy: A Guide to NumPy.'' USA: Trelgol
Publishing. \url{http://www.numpy.org/}.

\hypertarget{ref-scikit-learn}{}
Pedregosa, F., G. Varoquaux, A. Gramfort, V. Michel, B. Thirion, O.
Grisel, M. Blondel, et al. 2011. ``Scikit-Learn: Machine Learning in
Python.'' \emph{Journal of Machine Learning Research} 12: 2825--30.

\hypertarget{ref-peterson2009f2py}{}
Peterson, Pearu. 2009. ``F2PY; a Tool for Connecting Fortran and Python
Programs.'' \emph{Int. J. Comput. Sci. Eng.} 4 (4). Inderscience
Publishers, Geneva, SWITZERLAND: Inderscience Publishers: 296--305.
doi:\href{https://doi.org/10.1504/IJCSE.2009.029165}{10.1504/IJCSE.2009.029165}.

\hypertarget{ref-prim}{}
Prim, R. C. 1957. ``Shortest Connection Networks and Some
Generalizations.'' \emph{The Bell System Technical Journal}.
doi:\href{https://doi.org/10.1002/j.1538-7305.1957.tb01515.x}{10.1002/j.1538-7305.1957.tb01515.x}.

\hypertarget{ref-rainbolt:2016}{}
Rainbolt, Jessica Lovelace, and Michael Schmitt. 2017. ``The Use of
Minimal Spanning Trees in Particle Physics.'' \emph{JINST} 12 (02):
P02009.
doi:\href{https://doi.org/10.1088/1748-0221/12/02/P02009}{10.1088/1748-0221/12/02/P02009}.

\hypertarget{ref-szapudi1998}{}
Szapudi, István, and Alexander S. Szalay. 1998. ``A New Class of
Estimators for the N-Point Correlations'' 494 (1): L41--L44.
doi:\href{https://doi.org/10.1086/311146}{10.1086/311146}.

\end{document}